\documentstyle[11pt,aaspp4]{article}

\begin{document}

\received{}
\revised{}
\accepted{}

\lefthead{M. Catelan}
\righthead{Mergers of Globular Clusters}

\slugcomment{To appear in ApJ Letters}

\title{Color-Magnitude Diagrams of Merged Globular Clusters: 
       Metallicity Effects}

\author{M.~Catelan}

\affil{ NASA/Goddard Space Flight Center \\
        Laboratory for Astronomy and Solar Physics\\
        Code 681\\ 
        Greenbelt, MD 20771\\
        USA \\
        e-mail: catelan@achamp.gsfc.nasa.gov
      }

\begin{abstract}
Mergers of globular clusters (GCs) once associated with dwarf spheroidal (dSph) 
galaxies have recently been suggested as an explanation for the bimodal 
horizontal branches (HBs) of some Galactic GCs, most notably NGC 1851, 
NGC 2808, and NGC 6229. Through analysis of the available color-magnitude 
diagrams for the GCs in the Fornax and Sagittarius 
dSph satellites of the Milky Way, as well as their metallicity distributions, 
we argue that the merger of two GCs would most likely produce
a bimodal distribution in red giant branch (RGB) colors, or at least a
significant broadening of the RGB, due to the expected difference in metallicity
between the two merging globulars. No GC with a bimodal RGB is currently known,
and the tightness of the RGB sequences in the above bimodal-HB GCs
implies that a merger origin for their HB bimodality is unlikely.

\end{abstract}

\keywords{Stars: Hertzsprung-Russell (HR) diagram --- Stars:
          Horizontal-Branch --- Stars: Population II --- Globular Clusters:
          General --- Galaxies: Local Group --- Galaxies: Star Clusters.
         }

\clearpage

\section{Introduction}
The second-parameter phenomenon is a very intriguing anomaly affecting the
color-magnitude diagrams (CMDs) of Galactic globular clusters (GCs).
The morphology of the horizontal branch (HB) is observed to be 
primarily a function of the metallicity 
${\rm [Fe/H]} = \log\, ({\rm Fe/H})_{\rm GC} - \log\, ({\rm Fe/H})_{\sun}$, 
the ``first parameter."
However, superimposed on this main trend one finds intrinsic scatter,
caused by cluster-to-cluster variations in some unknown ``second parameter" 
(Sandage \& Wallerstein 1960; Sandage \& Wildey 1967; van den Bergh 1967). 

The idea that age is the second parameter has been very popular,
but evidence has been mounting suggesting a
much more complex scheme where several parameters may be simultaneously playing
more or less important r\^oles (see Stetson, VandenBerg, \&  Bolte 1996 and 
Fusi Pecci \& Bellazzini 1997 for recent reviews). One of the strongest 
arguments against age as the ``only"  second parameter 
is provided by the remarkable existence of the second-parameter effect within 
{\em individual} GCs (Rood et al. 1993), such as NGC 1851, NGC 2808, and 
NGC 6229 (Walker 1992b; Ferraro et al. 1990; Borissova et al. 1997). Since 
all stars within any individual GC presumably have the same age, a different 
second parameter must be responsible for the existence of both red and blue HBs 
within the same cluster.

In a recent paper, van den Bergh (1996, hereafter VDB96) has tentatively 
proposed that this argument against age as the second parameter
may be incorrect. In his new scenario, the merger between two 
GCs of different HB morphologies within dwarf spheroidal galaxies (dSph's) 
such as Fornax or Sagittarius could lead to bimodal HB types without violating 
the idea that age is the second parameter. Thus, the bimodal-HB clusters 
could have originated in mergers between second-parameter pairs of GCs in 
dSph's that were subsequently accreted by our Galaxy, as envisaged by
Searle \& Zinn (1978).

As noted by VDB96, the merger of two GCs could affect not only the HB
morphology but also other aspects of the CMD. The purpose of this 
{\em Letter} is to examine the implications of the merger scenario
for the morphology of the CMD, especially in the red giant branch
(RGB) region. We begin in the following section by presenting CMDs
produced by ``merging" GCs in the Sagittarius and Fornax dSph's and
in the LMC. We argue that these merged CMDs are not consistent with
the tight RGBs observed in Galactic GCs. In Sect. 3 we 
discuss the metallicity dispersions that one would expect within 
merged clusters. Finally, in Sect. 4 we summarize our results.

\section{Testing the Merger Hypothesis: GCs in Sagittarius, Fornax, and the LMC}

\subsection{GCs in the Sagittarius dSph}

As an example in support of the merger hypothesis, VDB96 has
pointed out that a merger involving the GCs Arp 2 (blue HB) and Terzan 7 (red 
HB) would lead to a bimodal HB. These two GCs, along with Terzan 8 and M54 
(NGC 6715), are associated with the Sagittarius dSph (Ibata, Gilmore, \& 
Irwin 1995; Da Costa \& Armandroff 1995; but see Vel\'azquez \& White 1995). 

%
\begin{figure}[t]                                                               
\figurenum{1}                                                                  
\plotfiddle{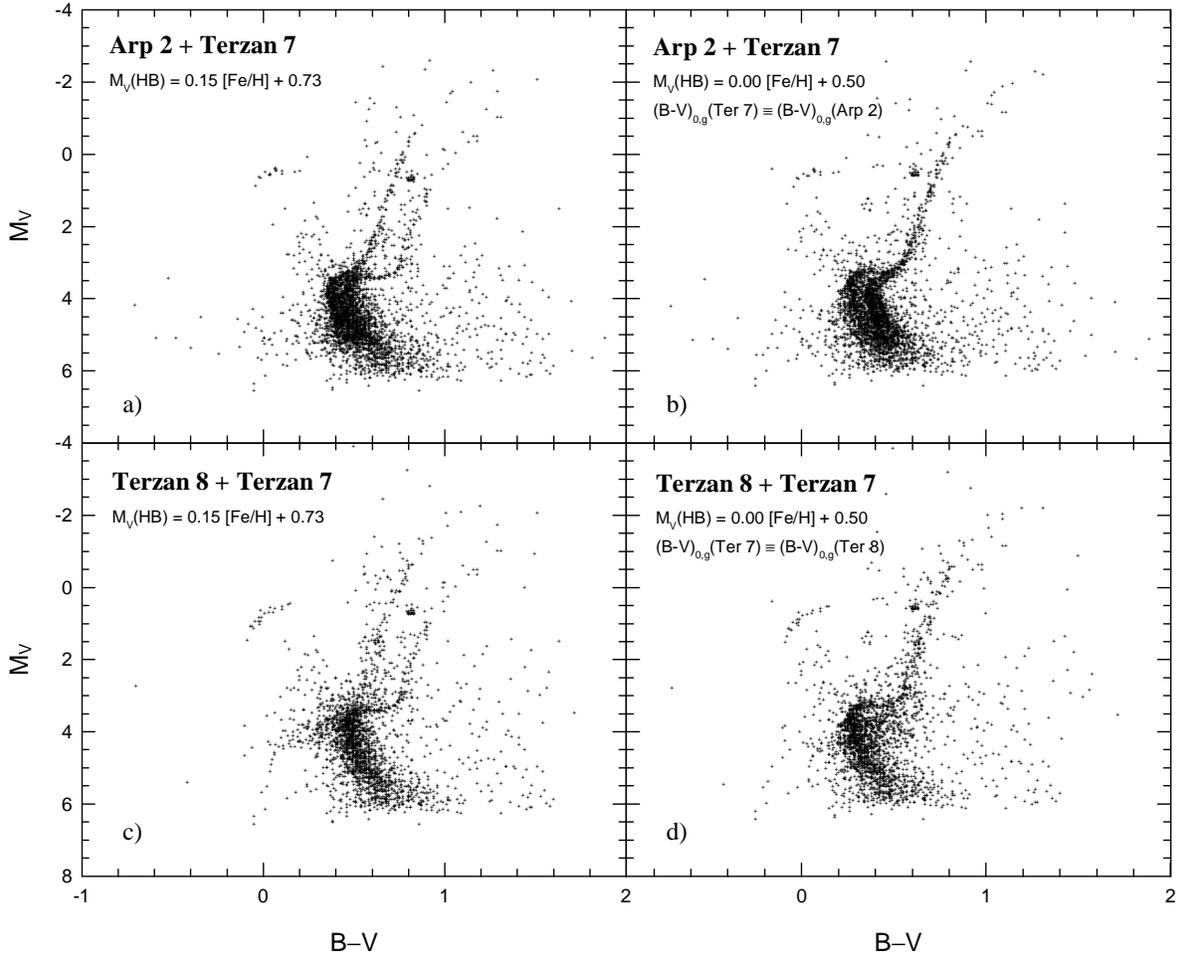}{4.65in}                                           
{0}{80}{80}{-248}{-235.0}  
\caption[]{\small Simulation of a ``merger" between Arp 2 and 
   Ter 7 (panel a)
   and between Ter 8 and Ter 7 (panel c) in the Sagittarius dSph. 
   Panels b and d are analogous to panels a and b, except that a match 
   is forced between the colors of the two RGBs at the HB luminosity
   level. The adopted $M_V({\rm HB}) - {\rm [Fe/H]}$ relations are also
   shown in the upper left-hand corners.}
\end{figure} 

In Fig. 1a, we combine the CMDs of Arp 2 and Ter 7 (Buonanno et al. 1995a,b)
to schematically
illustrate the implications of a merger between these two clusters. The 
colors and magnitudes were corrected for reddening according to the $E(\bv )$ 
values provided by Harris (1996), and the distance modulus determined adopting 
Walker's (1992a) distance scale. The choice of alternative distance scales
does not change the conclusions of this {\em Letter}. 

The most remarkable feature of the CMD shown in Fig. 1a, besides the bimodal 
HB, is the clearly bimodal RGB. This is due to the 
metallicity difference between these two clusters (e.g.,
Buonanno et al. 1995a,b): Ter 7 has a substantially redder RGB than does 
Arp 2, which is $\approx 1$ dex more metal-poor. In fact, inspection of Fig. 1a
reveals that almost the whole combined CMD presents signs of bimodality.
Such a clearly bimodal RGB distribution has never, to the best of our 
knowledge, been observed in a GC. 

In Fig. 1b, we arbitrarily ``slide" the Ter 7 CMD by 0.20 mag in $\bv$, 
so as to match the Arp 2 RGB color at the HB level. The two HBs are also 
aligned in brightness. As expected, the RGB bimodality has been substantially 
diminished, though not entirely---especially on the upper RGB. The bimodality 
in the subgiant branch and in the turnoff and main sequence are even more 
evident than in Fig. 1a. Note that, in producing Fig. 1b, we have effectively
assumed that the Ter 7 reddening value listed by Harris (1996), 
$E(\bv ) = 0.06$ mag, is too small by 0.2 mag. This, however, can
be safely ruled out (Buonanno et al. 1995b).

In Figs. 1c and 1d, we repeat the above experiment, but replacing Arp 2 with
Ter 8 (Ortolani \& Gratton 1990; Ortolani 1996). The situation is clearly
analogous to the Arp 2 plus Ter 7 case. The HB bimodality is again accompanied
by a very peculiar bimodal RGB.

An experiment involving the cluster M54 has not been possible, due to the 
multiple nature of the CMD in the direction of this cluster, caused (at least
in part) by contamination by Sagittarius field stars (Sarajedini \& Layden 
1995). However, since this cluster too is more metal-poor than Ter 7, 
we predict that a merger between M54 and Ter 7 would also lead 
to a bimodal RGB, although the large mass difference between the two clusters 
would make Ter 7's contribution to a combined CMD appear little more than a 
secondary contamination. It should be interesting to note, at any rate, that 
the composite nature of M54's CMD {\em may} be consistent with a merger 
event.

\subsection{GCs in the Fornax dSph}

The only other dSph associated with
the Galaxy known to contain GCs is Fornax, which has an anomalously high
specific frequency of GCs (Fusi Pecci 1987). Since Fornax is more 
distant, the CMDs for its GCs are not as well defined as in Sagittarius. 
Fortunately, however, the new Buonanno et al. (1996, hereafter BCFFPB96) 
data allow us to present preliminary simulations of the ``merger" between 
pairs of GCs in that remote dSph.

Fig. 1 in BCFFPB96 shows that Cluster 4 is possibly
the only one with a predominantly red HB in Fornax. The 
remaining GCs seem to have either intermediate or predominantly blue 
HB types (see also Beauchamp et al. 1995). The ``merger" between clusters 
1 and 4 in Fornax is illustrated in Fig. 2a. In Fig. 2b, a similar 
``merger" experiment involving clusters 3 and 4 is shown. Despite the
relatively poorly defined CMDs, the bimodality in the upper RGBs is 
still quite evident.

%
\begin{figure}[t]
\figurenum{2}                                                                  
\plotfiddle{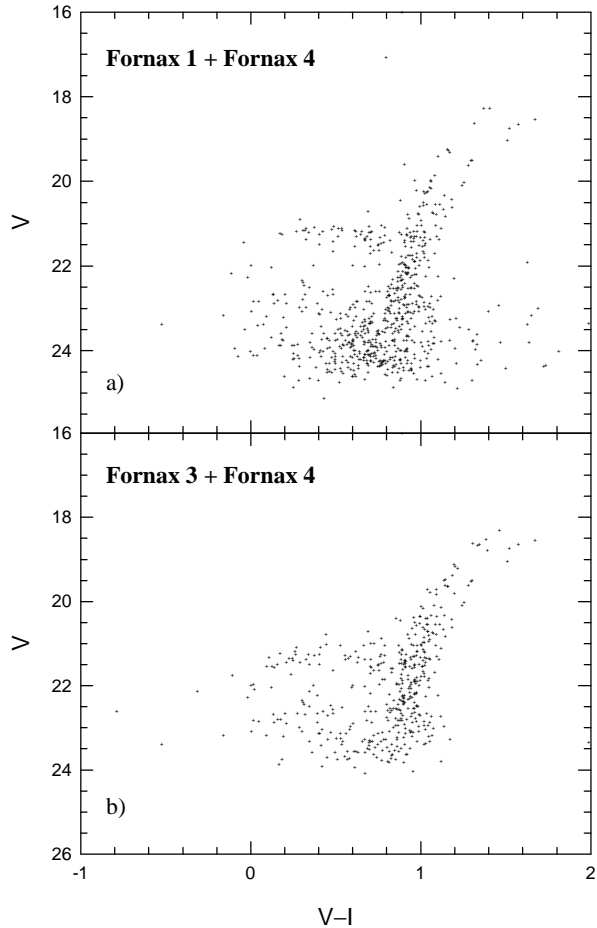}{4.5in}                                           
{0}{75}{75}{-225}{-135.0}  
 \caption[]{\small Same as Fig. 1, but for clusters 1 and 4 
   (panel a) and
   clusters 3 and 4 (panel b) in the Fornax dSph. The data were
   obtained from BCFFPB96. As in that study, 
   we have restricted the CMD of Cluster 3 (NGC 1049) to stars in
   the core of the cluster, and the CMD of Cluster 4 to an annulus 
   around the cluster center (cf. their Figs 1d and 1f, respectively).
            }
\end{figure}

VDB96 argues that Fornax clusters 1 and 2 may have
narrowly missed a merging event in the past. Even though a CMD for
Cluster 2 is not available for a straightforward comparison with
the Cluster 1 CMD of BCFFPB96, the
nominal metallicity difference between the two clusters, $\simeq 0.19$
dex (Buonanno et al. 1985; Dubath, Meylan, \& Mayor 1992), 
if real, would again show up as a bimodality in the merger's
RGB, if and when the merger product were finally captured
by the Galaxy.

\subsection{GCs in the LMC}

It has been suggested that some of the ancestral fragments that led to
the formation of the Galaxy in the Searle \& Zinn (1978) scenario were
actually significantly more massive than the dSph's, perhaps resembling
more closely the LMC in this respect (van den Bergh 1993). Indeed, some
of the Galactic GCs may have been stripped from the Magellanic Clouds
(van den Bergh 1994). Thus, we have also examined the Pop. II GCs in 
the LMC (Suntzeff et al. 1992) in connection with the VDB96 hypothesis.

Since none of the Pop. II GCs in the LMC has a red HB,
no pair of GCs whose merger would clearly lead to a bimodal HB could 
be found. Interestingly, mergers of {\em young} GCs do seem to be 
currently taking place in the LMC. However, these 
merging GCs were probably formed simultaneously within the same 
protocluster cloud, implying that the involved pairs of GCs have
essentially the same ages and chemical composition 
(Bhatia et al. 1991; Bica, Clari\'a, \& Dottori 1992; Rodrigues et al. 1994).
An exception seems to be provided by the pair NGC 1938/NGC 1939,
where the difference in age may be higher than 5 Gyr (Bica et al. 1992).
Apart from this case, no sign of bimodality would be currently present 
in the corresponding CMDs as a direct result of the mergers, if
these had taken place several Gyr ago. However, we speculate that an 
indirect influence {\em might} result due to the 
possible impact of the merger on the intracluster distribution in stellar 
rotational velocities and mass loss rates. 

\section{Metallicity Distributions}

In the previous section, we have found that the mergers of GCs tend to produce
bimodal RGB distributions. The reason for this, as already discussed, 
is the difference in metallicity
between the merging GCs. That the RGB color should become redder with 
increasing metallicity has been known observationally since Sandage \& 
Wallerstein (1960) and Wildey (1961). The effect is also well understood 
theoretically (Hoyle \& Schwarzschild 1955). Moreover, the shape of the RGB 
also depends strongly on [Fe/H], as recently discussed by 
Da Costa \& Armandroff (1990) and Ortolani, Barbuy, \& Bica (1991). 

Two questions arise from this: how tightly is the [Fe/H] dispersion
constrained within {\em single} GCs, especially those with bimodal HBs?
And what is the expected dispersion in [Fe/H] in the merger scenario,
given the observed metallicity dispersion of the GC systems in the dSph
galaxies?

As to the first question, it is important to note that current upper limits 
on the metallicity dispersions within individual GCs are {\em extremely} tight.
In fact, as stated by Suntzeff (1993), ``0.04 dex is a realistic upper
limit to the average cluster metallicity inhomogeneity." 
Part of the reason for this remarkable accuracy is exactly the fact 
that the RGB color and shape depend so sensitively on [Fe/H].  
As far as the {\em bimodal-HB clusters} are concerned,
NGC 1851 has been studied by Da Costa \& Armandroff (1990). According
to their analysis, the $3\sigma$ upper limit on the internal metallicity
dispersion is 0.07 dex. The CMD study of NGC 2808 by Fusi Pecci \&
Ferraro (1996) similarly implies a $3\sigma$ upper limit of $\approx 0.15$
dex for the dispersion in [Fe/H] within the cluster. For NGC 6229,
Borissova et al. (1997) have estimated that the width of the RGB is
essentially consistent with their measurement errors, implying no
intrinsic dispersion in [Fe/H] within the cluster. In fact, among all 
the GCs for which CMDs are available, $\omega$Cen (NGC 5139), M22 
(NGC 6656), M92 (NGC 6341), and M54 (in Sagittarius) are the only 
cases for which an internal metallicity dispersion has been proven 
or suspected (Suntzeff 1993 and references therein; Sarajedini \& Layden 
1995). 

Using the latest metallicity and CMD data available in the literature,
we have performed a detailed analysis of the [Fe/H] and HB 
morphology distributions of the GCs in the Fornax and Sagittarius 
dSph's, as well as in the LMC and in the Fusi Pecci et al. (1995) 
groups of GCs that may have shared a possible common 
origin within a Searle \& Zinn (1978) ``building block" of the Galaxy 
(see also Lynden-Bell \& Lynden-Bell 1995). Our results indicate that 
{\em a general prediction of the merger
hypothesis is that many mergers involving pairs of globulars
with different metallicities would be expected for every merger 
involving a pair of GCs with different HB morphology but
essentially the same metallicity}. By ``different metallicities" 
is meant $\Delta {\rm [Fe/H]} > 0.04$ dex (Suntzeff 1993).
Searle (1977) has also pointed out that a merger might lead to an               
inhomogeneous, and possibly bimodal, metallicity distribution. 
In fact, a bimodal HB would be much more likely to
result from the merger between two GCs of different metallicities than
between two clusters of nearly the same metallicity. This is simply a
consequence of the fact that metallicity is the {\em first} parameter.
At the same time, bimodal RGBs are a much more
natural consequence of the mergers of GCs than are bimodal HBs, since
{\em the impact of even tiny differences in [Fe/H] on stellar 
colors are greatly amplified during the RGB phase}.

These conclusions are also supported by an analysis of the GC metallicity 
distribution in NGC 147, NGC 185, and (to a lesser extent)
NGC 205---the dSph companions to M31
known to contain GCs (see, e.g., Minniti, Meylan, \& Kissler-Patig 1996);
and by the metallicity distributions in the M31 GC ``groups" 
proposed by Ashman \& Bird (1993).

\section{Conclusions}

The above analysis indicates that, in all likelihood, mergers of GCs cannot 
have been responsible for the HB bimodality in NGC 1851, NGC 2808, and 
NGC 6229, since the size of the metallicity dispersion within these clusters
is very tightly constrained, as opposed to what would be expected in the
merger scenario. Moreover, bimodal RGBs and RGBs which are 
substantially wider than those which are usually observed would be much more 
frequent in the Galactic GC system as a whole, if mergers of bona-fide old GCs 
were responsible for the existence 
of bimodal-HB clusters. We suggest that signs of merger events between old GCs 
should be searched for primarily in the RGB region of the CMD, where 
even small metallicity differences between the merging globulars are expected 
to greatly affect the resulting distributions.

\acknowledgments

The author is indebted to Dr. S. Ortolani, who has kindly provided the
new reduced Ter 8 data employed in Fig. 1; and to Dr. M. Bellazzini and
Dr. F. Fusi Pecci for providing their ESO-NTT data for the Fornax GCs prior
to publication. The author is grateful as well to Dr. E. Bica and 
Dr. L. Girardi for providing useful information, and to 
Dr. A. V. Sweigart for his continuous encouragement and useful discussions.
This research was supported in part by NASA grant NAG5-3028. This work was 
performed while the author held a National Research Council--NASA/GSFC Research 
Associateship.

\end{document}